\documentclass[aps,prl,twocolumn,showpacs,superscriptaddress]{revtex4-1}

\usepackage{rotating, graphicx}
\usepackage[utf8]{inputenc}
\usepackage{caption}
\usepackage{siunitx}
\usepackage{amsmath}
\usepackage{gensymb}
\usepackage{amssymb}
\usepackage{enumerate}
\usepackage{color}
\usepackage{hyperref}
\usepackage[american]{babel}
\usepackage{subcaption}
\usepackage{etoolbox}
\usepackage{floatrow}
\usepackage{nicefrac}
\usepackage{enumitem}

\patchcmd{\thebibliography}{\section*{\refname}}{}{}{}

\newcommand{\fig}[1] {Fig.~\ref{fig:#1}}
\newcommand{\tab}[1] {Tab.~\ref{tab:#1}}

\begin{document}

\title{\textit{PeopleTraffic}: a common framework for harmonizing privacy and epidemic risks}

\newcommand{\corresponding}[1]{\altaffiliation{#1}}

\newcommand{\afftifpa}[0]{\affiliation{TIFPA/INFN Trento, via Sommarive 14, 38123~Povo, Trento, Italy}}
\newcommand{\affmi}[0]{\affiliation{Università Statale di Milano e Fondazione RCM, via Celoria 18, Milano, Italy}}
\newcommand{\affcern}[0]{\affiliation{Physics Department, CERN, 1211~Geneva~23, Switzerland}}
\newcommand{\affrulex}[0]{\affiliation{Rulex Innovation Labs, via Felice Romani 9, 16122~Genova, Italy}}

\author{Ruggero~Caravita}
\corresponding{ruggero.caravita@cern.ch}
\afftifpa{}




\date{\today}

\begin{abstract}
PeopleTraffic is a proposed initiative to develop a real-time, open-data population density mapping tool open to public institutions, private companies and the civil society, providing a common framework for infection spreading prevention. The system is based on a real-time people' locations gathering and mapping system from available 2G, 3G and 4G mobile networks operators, enforcing privacy-by-design through the adoption of an innovative data anonymizing algorithm inspired by quantum information de-localizing processes. Besides being originally targeted to help balancing social distancing regulations during the Phase-2 of the COVID-19 pandemics, PeopleTraffic would be beneficial for any infection spreading prevention event, e.g. supporting policy-makers in strategic decision-making.
\end{abstract}

\maketitle{}

\section{Introduction}

The adoption of social distancing regulations, enforcing physical separation between individuals to limit population density and reducing individual-to-individual contacts, has been so far the most effective tool employed by national countries to mitigate the COVID-19 pandemics cases' shock on their national sanitary systems \cite{lancet_socdist:20}.
The effectiveness of these regulations was predicted by most epidemic models \cite{nature_modelling:20} based on the hypothesis that individual-to-individual contact is the main transmission process driving infection spreading (see \cite{setti_airborne:20} for a review). Their strong corroboration by the observations lead to conclude that local population density in the proximity of an individual has a strong a-priori predictive power on the likeliness of that individual to get infected.
\par
Civil society's response to these strict social distancing regulations and restrictions has been widely positive (i.e. with a high degree of acceptance) in most countries involved in  the first phase of the COVID-19 pandemic (e.g. $\sim\,85 \%$ for the Italian case, see \cite{pepe_isi:20}). This indicates that most people act according to regulations for their own and common interests, and also suggests that wide adoption of positive individual behaviours is likely if all citizens are given a set of good practices and tools to follow them. 
\par
It became also clear, on the other hand, that strict social distancing regulations have profound impacts on countries' economies \cite{imf_weo:20}, and their partial to full relaxation becomes necessary to mitigate the adverse economical effects as soon as pressure on the sanitary systems is sufficiently relaxed. This marks the beginning of the co-existence period with the infection (the so-called pandemic Phase-2), which is expected to last approximately until a vaccination becomes available to the mass public (estimated mid-2021 for the COVID-19 case). 
\par
The main risk connected to relaxing social distancing measures is to increase the risk of epidemic resumption due to increased people individual-to-individual contacts. The possibility to differentiate, rather than relieve, social distancing policies in different contexts is likely to mitigate that risk and consequent novel shocks on national sanitary systems. In what forms this differentiation can take place, it is one of the most discussed themes at the moment by policy-makers of most countries involved in the epidemic, as the ability to efficiently tune social distancing rules in economic, social and personal life contexts (and rationally distribute the diffusion risk) could have a profound impact on each country economic performance during Phase-2.
\par
Here we propose an initiative to develop a real-time, open-data population density and flux mapping system, hypothetically named \textit{PeopleTraffic}, in an analogy to the well-known Google Traffic system for road traffic. The proposed system enforces privacy-by-design and allows privacy levels to be externally and transparently regulated through the use of an innovative anonymizing algorithm specifically designed for mapping applications and inspired by quantum information de-localizing processes. This tool relies on GSM, not bluetooth sensing. Contrary to many other systems being developed that are based on direct proximity sensing via Bluetooth, and subsequent notifications in case of proximity with a subsequently identified infected individual, this scheme relies on determining the average local area person density via GSM/UMTS/LTE. It is meant to enable a win-win reduction of pandemic diffusion risks linked to people movements by supporting their autonomous risk prevention evaluations and decisions with the knowledge of real-time population density and fluxes. Furthermore, it would allow social distancing to be adapted to a variety of contexts, providing valuable quasi real-time information to individuals and policy-makers in taking decisions (both at the individual and strategic levels) based on a clear view of people density, which has proved so far one of the most reliable a-priori predictors of infection spreading. Finally, in the long-term, it could be a driver of innovation and economic activity, for instance as a tool for accurate flux and market analyses and business cases identification.
\par
In order for the population density and flux knowledge to be effective for individual-to-individual contact prevention, spatial and time resolutions close to the individual proximity (i.e. tens of meters and minutes) are necessary. This is key to allow virtuous behaviour orienting at the individual scale, like choosing the right means of transportation for commuting, or the right places and timing for necessary and leisure activities. This requires however the system to deal with individual privacy by design, preventing especially the risk of individual re-identification and tracing. 
\par
Bringing the knowledge of real-time population density and flux to everyone thus necessitates solving two main challenges: how to technically produce maps of sufficient resolution to be useful for individual-to-individual contact prevention, and a strong privacy risks mitigation approach enforcing full transparency and privacy-by-design to the data handling and sharing system.
\par
In this work, we review the possibilities offered by present-day technologies in constructing real-time population density and flux maps through the use of existing Real-Time Locating System (RTLS) with sufficient spatial and temporal resolutions to get close the individual scale. Secondly, we approach the privacy concern in quantitative manner, i.e. describing a tunable algorithm providing privacy-by-design by preventing individuals' localization at the level of the each data provider. Different levels of privacy could be transparently set (at the price of statistics necessary to reach a certain level of resolution) for effective policing to be allowed. Last, we lay out the design of the information processing system, highlighting its principal implementation characteristics.

\section{Measuring people locations in real time}

The main positional data acquisition methods we considered here are these that can be obtained from real-time data analysis of nowadays capillary mobile networks (i.e. these adopting GSM, UMTS, and LTE network technologies), i.e. not requiring any software to be installed on individual user equipment (UE).
\par
Modern mobile networks, from the physical layer point of view, allow many of such techniques with varying accuracies and required computing power \cite{vin_overview:15}. What method can be implemented on each territory is thus mainly in the hands of each mobile network provider capabilities and data sharing policies. It's reasonable to assume that different accuracies would be obtained from different providers each adopting its own localization approach from the simplest to the most elaborated method demonstrated so far. 
\par
Here we consider the two limiting cases: A) the mere counting of number of connections per mobile network cell (also known as localization through cell identity, CI) and B) accurate individual UE positioning by the network via the highest accuracy available methods, e.g. RSSI triangulation or observed time difference of arrival (OTDOA). 
\par
Scenario A, i.e. counting the number of connections per network cell, has the advantage to be applicable to most network technologies, including public WiFi hotspots. It is the simplest and most anonymous approach, based on data that most network providers already possess (and even sell for market analytics purposes in some cases \cite{nyt_fcc:20}). Its mapping accuracy is however limited by the size of deployed network cells. 
\par
Scenario B, i.e. measuring independently each UE position, can provide sub-cell resolution at the price of higher requirements in terms of complexity, data processing power and personal data handling. Most network providers already developed all necessary infrastructures during the LTE technology deployment phases to meet regulatory emergency call positioning requirements. Indeed, the importance of accurately geo-locating emergency phone calls was recognized to be so socially impacting already that it was made part of the LTE network standard: as of 2020, 70 \% of emergency calls in the United States are positionally located within \SI{50}{\meter} \cite{ryden_ltepos2:16}. Individual UE locating methods usually require a data anonymizing step after mapping is performed to comply to General Data Protection Regulations (see Privacy section). Mapping and anonymizing algorithms have typically to be run by the network operators to avoid distributing non-anonymous data. As for the former case, achievable resolutions increase proportionally to mutual inter-distances between transceiver stations.

\subsection{Distribution of mobile network cells}

Expected positional accuracies of both scenario A and B are proportional to local densities of mobile network cells. Thus, mapping them on country scales is necessary to evaluate the realizability of a mass RTLS as the one here discussed in a realistic use-case scenario. 
\par
We conducted a survey of existing cell locations and sizes to determine the granularities of GSM, UMTS and LTE networks and estimate the positional accuracies that could be obtained by network operators adopting the two approaches discussed above.
\par
\begin{figure*}[h!tbp]
	\centering		
    \includegraphics[width=0.25\linewidth]{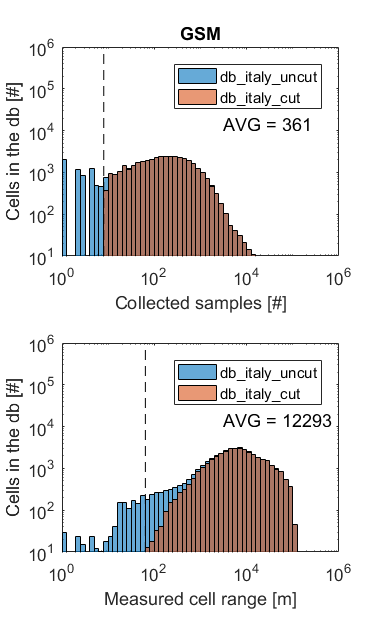}	
    \includegraphics[width=0.25\linewidth]{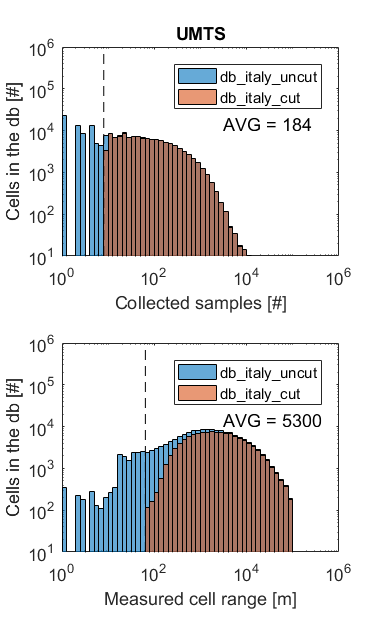}	
    \includegraphics[width=0.25\linewidth]{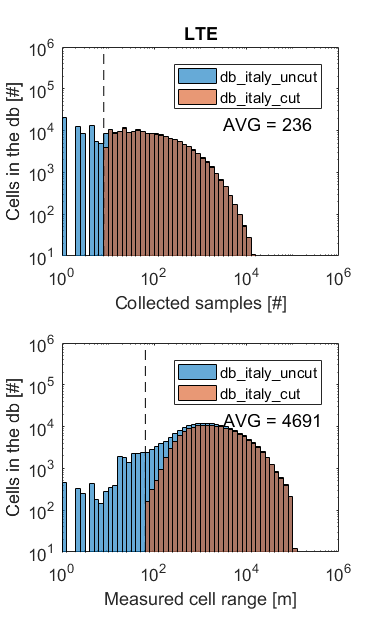}	    
	\caption{Overview of the Mozilla Location Service database: raw (blue) and data quality cut (brown) distributions of single cells' number of geo-tagged samples (top row) and measured ranges (bottom row) for GSM (left), UMTS (center) and LTE (right) technologies on the Italian territory.}	
	\label{fig:db_overview}
\end{figure*}

The study was conducted by merging all available cells identification codes and geo-locations from the OpenCellID \cite{open_cell:20} and Mozilla Location Services \cite{mozilla_locserv:20} free databases. Each cell is uniquely identified by its Mobile Country Code (MCC), Mobile Network Code (MNC), Location Area Code (LAC) and Cell Identification number (CID). It's worth summarizing, for the following discussions, how  these databases are constructed. They are built upon big datasets of geo-tagged mobile connections to individual cells of the networks from physical people mobile phones participating in the mapping initiatives, logging and streaming their anonymous GPS positions and connected network cells identification numbers while moving. Each set of mobile phones' logged GPS positions is then clustered on a per-cell basis, from which the centroid position is calculated by averaging all samples' positions. Cell connection ranges are also estimated by the spatial spread of each cluster points. The number of samples available per cell is widely distributed due to this statistical acquisition modality, spanning from the $ < 10 $ samples of infrequently visited cells up to the $ > 10000 $ samples of most frequently visited cells (\fig{db_overview}, top row). Ranges are similarly widely distributed, spanning from a few \si{\meter} to \SI{100}{\kilo\meter} (\fig{db_overview}, bottom row). We applied a data quality cut to the dataset rejecting all these cells with $ < 8 $ samples, giving rise to an nonphysical tail in the ranges distribution $ < \SI{50}{\meter} $ (\fig{db_overview}, dashed lines). It is worth noting also that, despite the notable dimensions of these location databases, their statistical acquisition modality can give rise to systematic errors when comparing different contexts. A first issue is spatial non-uniformity in sampling/partial territorial coverage. This issue was mitigated analyzing territories with comparable amounts of acquired samples per unit surface (given in the databases). A second concern is the completeness of the databases for small or local mobile network providers in terms of actual active cells on the territory. For this reason, the analyses reported here are given separately per network provider and limiting to those with major shares of physical-people Sim cards in Italy: TIM, Vodafone and Wind Tre, covering 30.0 \%, 27.2 \% and 24.7 \% market respectively \cite{agcom_report:19}.

\begin{figure*}[h!tbp]
	\centering		
    \includegraphics[width=0.8 \linewidth]{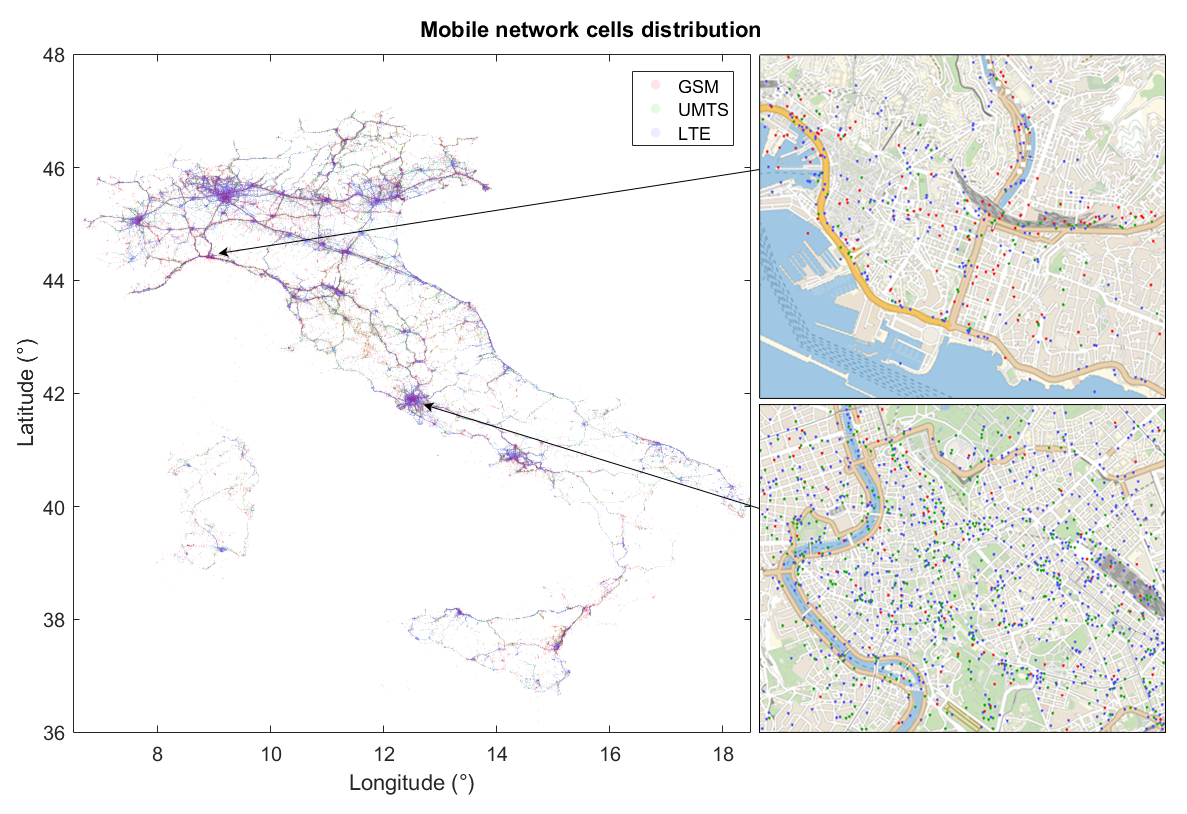}		
	\caption{Spatial distribution of mobile network cell positions from all network operators on the whole Italian territory (left) and in typical Italian urban contexts: Rome (right bottom) and Genoa (right top) historical city centers.}
	\label{fig:italy_dist}
\end{figure*}

By this method, about $ 1.25 \cdot 10^6 $ unique mobile cells were listed on the Italian territory (see \fig{italy_dist}, showing with different color the density distribution of GSM, UMTS an LTE cells). The list of cells was subsequently reduced to the scale of single urban territories. Here we considered the two historical city centers of the cities of Rome, a very large-sized town (\fig{italy_dist}, right bottom), and of Genoa, a large-sized town (\fig{italy_dist}, right top). The former, being the reference case-study city used in \cite{ratti_rturban:10}, allows also direct spatial resolution comparisons between the methods, while the latter is an example of large-sized town with complex orography (seaside, rivers and hills) as typically found in Italian provinces. 

In order to compare GSM, UMTS and LTE network cell sizes, one has first to determine their expected spatial shapes (here referred to as \textit{connection domains}) to be  able to then reconstruct their size distributions. Connection domains are here defined as those surface domains in whose boundaries single mobile devices can connect with a certain probability to that particular network cell. Constructing connection domains associated to each cell thus implies accounting for the typical behaviour of mobile network devices to connect to their closest cell, when any is available, starting from the cell positions $ \vec{p}_i $ and ranges $ r_i $ and evaluating any point in the surface. Solving this problem in the general case is a genuine challenge, suggesting the use of a Monte-Carlo approach. Here we adopted an approximated approach valid in the high network cell density limit. In this regime, cell inter-distances are much smaller than the maximum connection ranges allowed by the technology and observed ranges are set almost entirely by the transceiver positions $ \vec{p}_i $ (thus neglecting $ r_i $). Also, it implies that the surface is completely covered by connection domains, forming a tessellation set by $ \vec{p}_i $. This type of arbitrary tessellations constructed upon a set of generating points $ \vec{p}_i $ are usually known as Voronoi tessellations \cite{fortune_voronoi:95}.
\par
Voronoi tessellations are constructed from a set of generating positions $ \vec{p}_i $ as follows: the domain relative to the $i$-th point is defined by all points $ \vec{x}_j $ of the surface for which $ \vec{p}_i $ is the closest generating point, according to the distance function $d(\vec{x}_j,\vec{p}_i)$, here set to the two-dimensional Euclidean distance $d(\vec{x}_j,\vec{p}_i) := || \vec{x}_j - \vec{p}_i ||$. At first glance, it may look like the use of Voronoi tessellations with the Euclidean distance for constructing connection domains is only grossly approximating the real process of mobile devices connecting to their most intense nearby transceiver. One might argue that this process implicitly assumes equal transmission power and reception sensitivity for all cell transceivers and uniform antenna angular emission patterns, and that nearby cells overlap each other to allow handovers. 
\par
The first criticism would be correct if \textit{exact} cell towers positions would be used for constructing the connection domains. Here, however, cell positions $ \vec{x}_i $ taken from location databases are calculated as cluster centroids from the original geo-tagged user positions. Voronoi tessellations with the Euclidean distance are nothing but the graphical representations of the spatial domains associated to the clusters calculated by the well-known K-means clustering algorithm. Indeed, K-means clusters are obtained minimizing the sum-of-squares of all Euclidean distances between N generating points (the cluster centroids) and the dataset samples, which is the same process to build a Voronoi tessellation with the Euclidean distance.
\par
The second criticism is correct in that it possible for a device to connect to a cell even on the outside of its connection domain. However, this is unlikely to happen frequently, as normally, when a mobile device leaves a cell connection domain, automatic handover occurs transferring the connection to the nearby cell with its better signal reception. For this to occur nonetheless, either the device has to move at high velocity or channel saturation has to occur, i.e. nearby cells handover connections having all connection channels used. 
\par

\begin{figure*}[h!tbp]
	\centering		
    \includegraphics[width=0.8\linewidth]{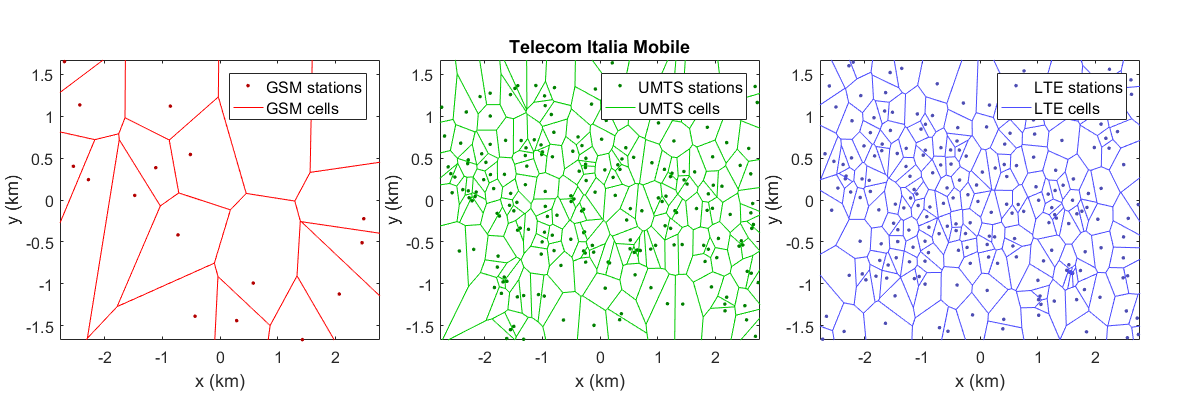}
    \includegraphics[width=0.8\linewidth]{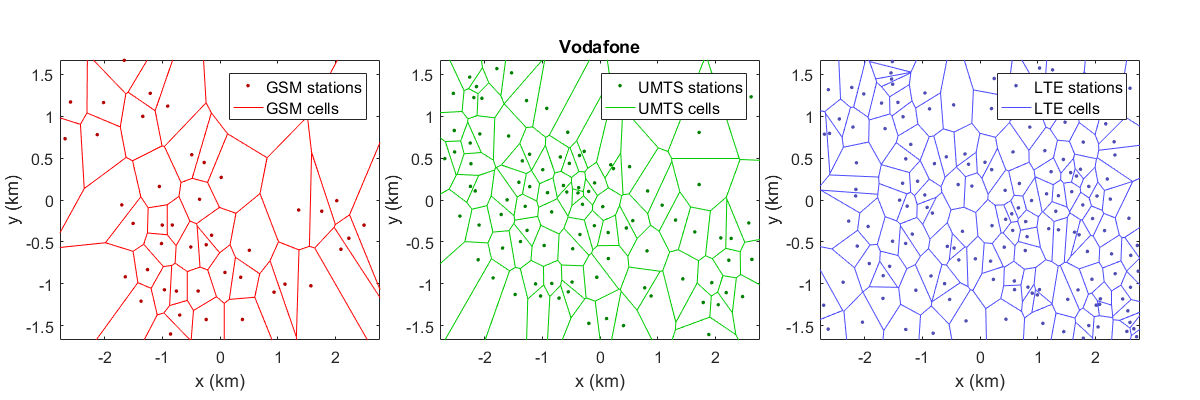}
    \includegraphics[width=0.8\linewidth]{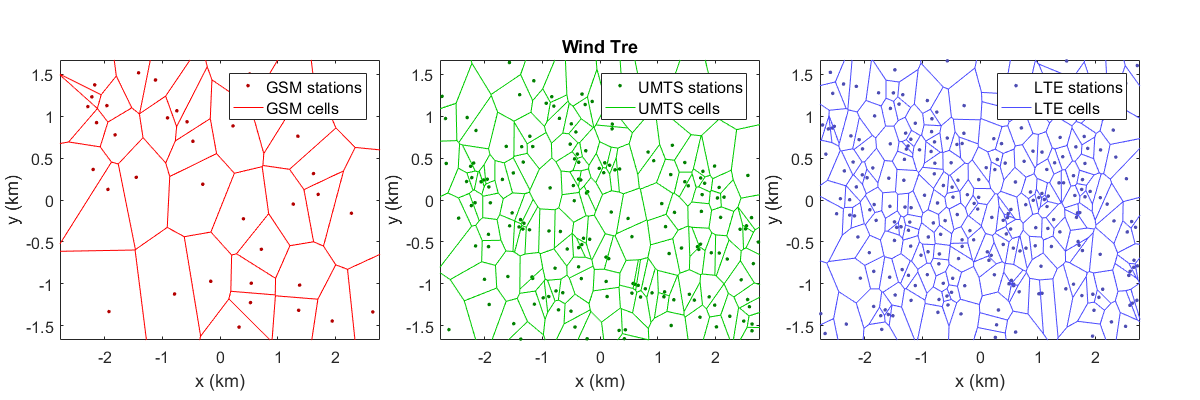}    
	\caption{Connection domains spatial chart (see text) for the TIM (top row), Vodafone (middle row) and Wind Tre (bottom row) network operators and for the GSM (left column), UMTS (middle column) and LTE (right column) technologies in a very large-sized town (here the historical center of the city of Rome).}
	\label{fig:rome_voronoi}
\end{figure*}

\begin{figure*}[h!tbp]
	\centering		
    \includegraphics[width=0.8\linewidth]{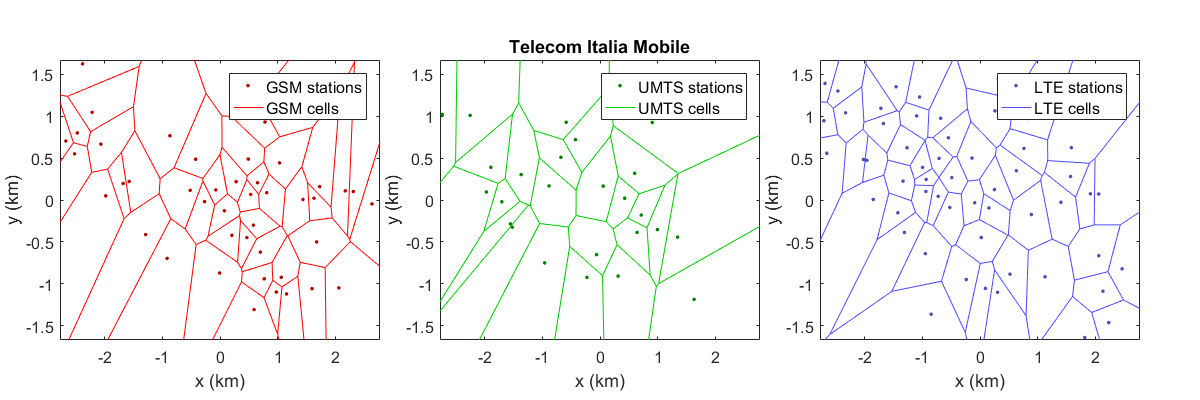}
    \includegraphics[width=0.8\linewidth]{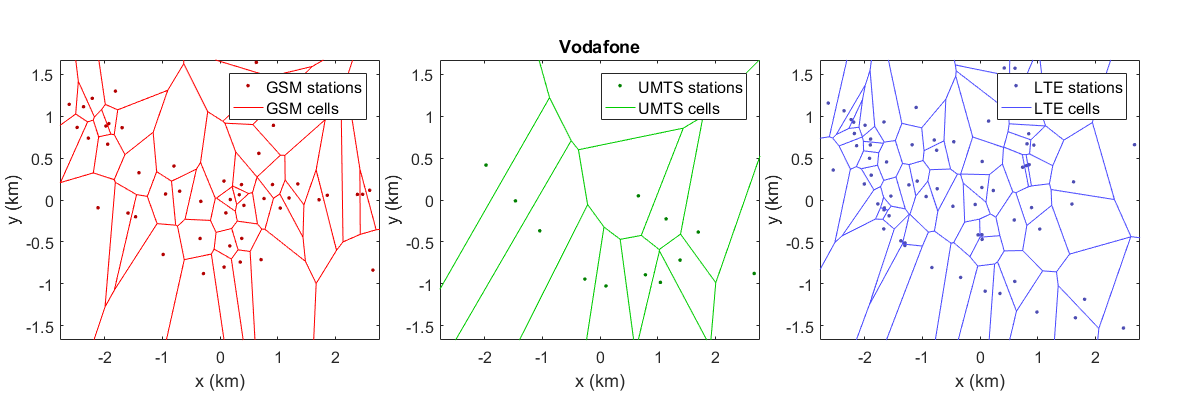}	   
    \includegraphics[width=0.8\linewidth]{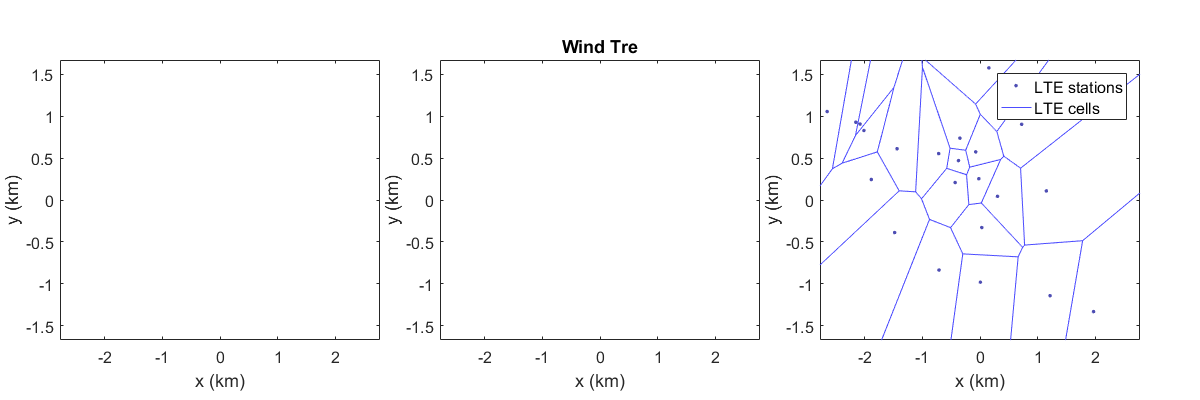}	      
	\caption{Connection domains spatial chart (see text) for the TIM (top row), Vodafone (middle row) and Wind Tre (bottom row) network operators and for the GSM (left column), UMTS (middle column) and LTE (right column) technologies in a large-sized town (here the historical center of the city of Genoa).}
	\label{fig:genova_voronoi}
\end{figure*}

\begin{figure*}[h!tbp]
	\centering		
	\includegraphics[width=0.25\linewidth]{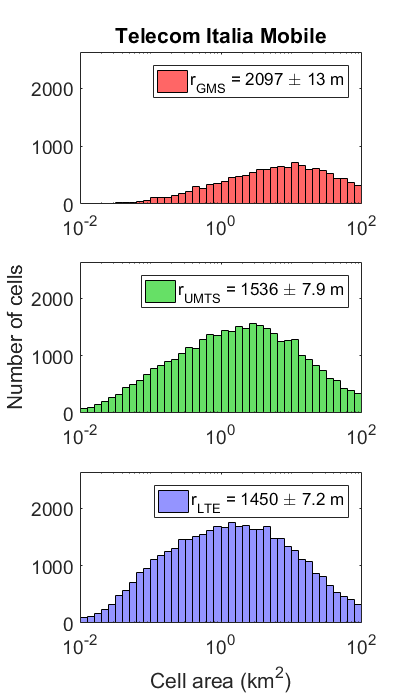}	
	\includegraphics[width=0.25\linewidth]{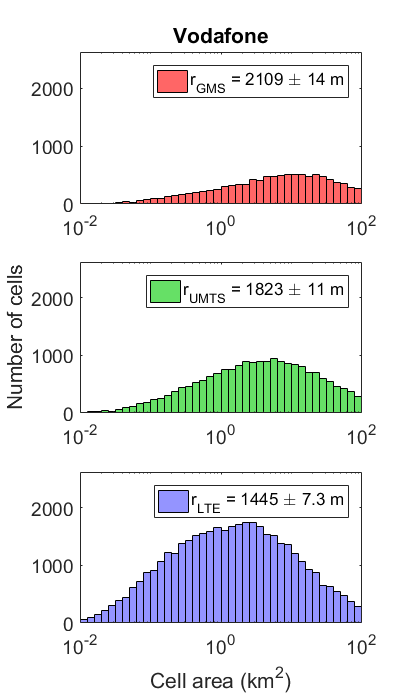}
	\includegraphics[width=0.25\linewidth]{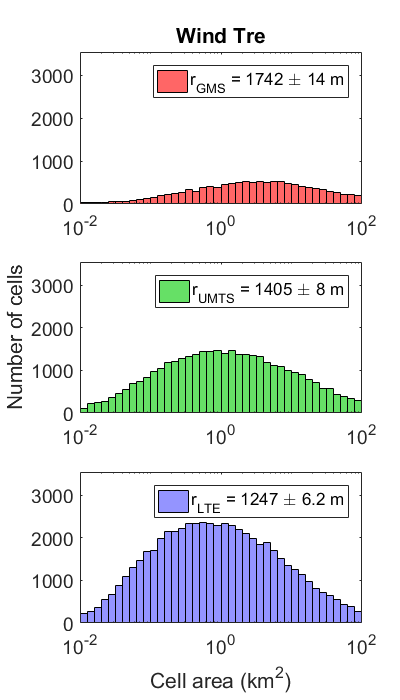}
	\caption{Distribution of GSM (red), UMTS (green) and LTE (blue) cell surfaces (cut at \SI{100}{\kilo\meter\squared}) for the three principal network operators on the whole Italian territory.}
	\label{fig:italy_hist}
\end{figure*}

\begin{figure*}[h!tbp]
	\centering		
    \includegraphics[width=0.25\linewidth]{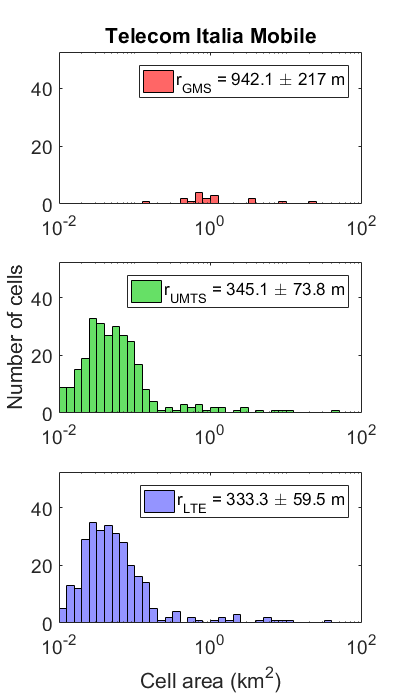}	
    \includegraphics[width=0.25\linewidth]{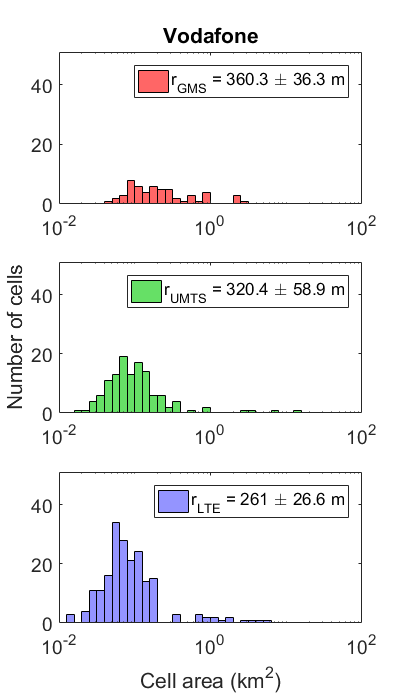}  
    \includegraphics[width=0.25\linewidth]{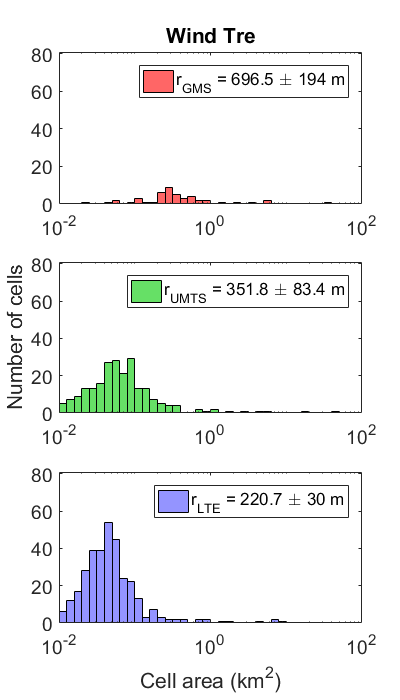}
	\caption{Distribution of GSM (red), UMTS (green) and LTE (blue) cell surfaces (cut at \SI{100}{\kilo\meter\squared}) for the three principal network operators in a very large-sized town (here the historical center of the city of Rome).}
	\label{fig:rome_hist}
\end{figure*}

\begin{figure*}[h!tbp]
	\centering		
    \includegraphics[width=0.25\linewidth]{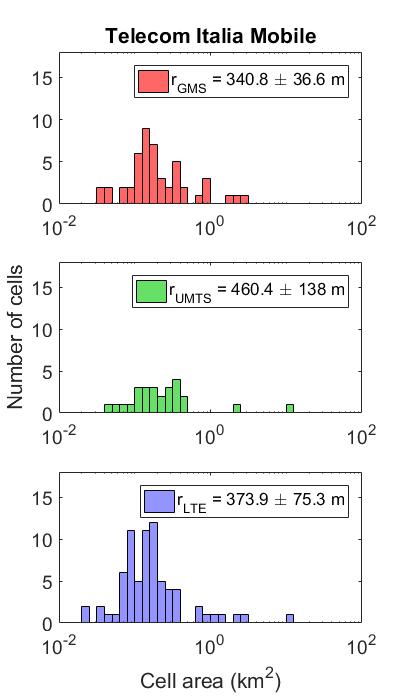}	
    \includegraphics[width=0.25\linewidth]{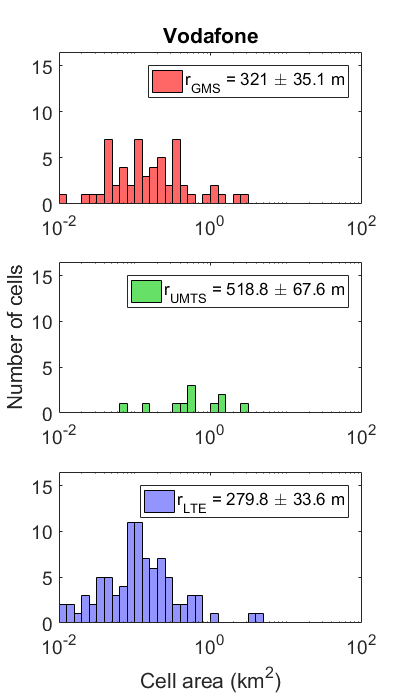}
    \includegraphics[width=0.25\linewidth]{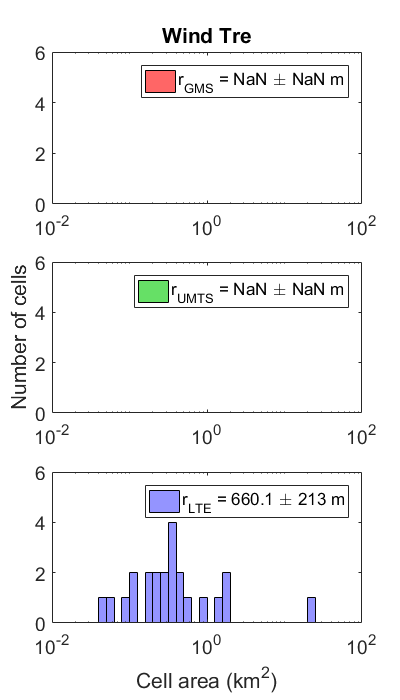}    
	\caption{Distribution of GSM (red), UMTS (green) and LTE (blue) cell surfaces (cut at \SI{100}{\kilo\meter\squared}) for the three principal network operators in a large-sized town (here the historical center of the city of Genoa).}
	\label{fig:genova_hist}
\end{figure*}

The distributions of connection domains for the TIM (MNC-1), Vodafone (MNC-10) and Wind Tre (MNC-88) network operators were constructed using the subset of cells' position associated to each MNC code. Their spatial distribution for the whole Italian territory are omitted due to their complexity, while they are shown for the Rome and Genova historical city centers in \fig{rome_voronoi} and \fig{genova_voronoi} respectively. 
The statistical analysis of their surfaces are shown in \fig{italy_hist} for the whole Italian territory and in \fig{rome_hist}, \fig{genova_hist} for the historical city centers of Rome and Genoa respectively. Associated network cell radii $ r_\mathrm{GSM} $, $ r_\mathrm{UMTS} $ and $ r_\mathrm{LTE} $ (also reported in the figures) were calculated under the approximation of average circular cell (i.e. inverting $ S = \pi r^2 $), valid in the high number of cells limit.
\par
The distributions on the whole Italian territory show that deployed GSM, UMTS and LTE networks have increasingly higher spatial granularities as expected, with average cell radii of $ \approx \SI{2.0}{\kilo\meter} $, $ \approx \SI{1.6}{\kilo\meter} $ and $ \approx \SI{1.4}{\kilo\meter} $ respectively and $ ~10\% $ fluctuations from network operator to network operator. Average cell radii dramatically reduce while zooming on a single urban context. In the case of Rome, a very large-sized town, these reduce to $ \approx \SI{0.7}{\kilo\meter} $, $ \approx \SI{0.4}{\kilo\meter} $ and $ \approx \SI{0.3}{\kilo\meter} $ whereas in the case of Genoa, a large-sized town, they similarly reduce to $ \approx \SI{0.3}{\kilo\meter} $, $ \approx \SI{0.5}{\kilo\meter} $ and $ \approx \SI{0.4}{\kilo\meter} $. 

\subsection{LTE networks positional accuracies}

An accurate estimate of positional accuracy of deployed LTE networks can now be obtained by combining the measured cell radii with an evaluation of the intrinsic resolution of CI and OTDOA localization techniques in a controlled test scenario including noise and three-dimensional effects. Here we considered the horizontal accuracy analyses reported in Ryd\'en \textit{et al.} \cite{ryden_ltepos:15}, which established that spatial resolutions of $ \sigma^\mathrm{A}_\mathrm{LTE} = \SI{268}{\meter} $ and $ \sigma^\mathrm{B}_\mathrm{LTE} = \SI{27}{\meter} $ (Gaussian standard deviations at 68\% error) can be obtained for scenario A and B respectively with regularly-spaced outdoor cells of \SI{167}{\meter} radius (i.e. 1/3 of the interdistance between transceivers considered in the study). The analysis of \cite{ryden_ltepos:15}, in other words, reports that indirect UE position detection by cell identity has a resolution (at 68\% error level) $ \sigma^\mathrm{A}_\mathrm{LTE} \approx 1.6 \cdot r_\mathrm{LTE} $. 
\par
These resolutions were scaled by the ratio of the cells radii to get the estimated ground resolutions of deployed networks. A summary of the obtained positioning resolutions for the different contexts is given in \tab{lte_accuracies}. By using anonymous CI data only and combining data of different LTE network providers, resolutions up to $ \sim \SI{250}{\meter} $ can be obtained in dense urban areas. Using OTDOA, a factor of 10 in resolution is gained, bringing it down to $ \sim \SI{25}{\meter} $.

\begin{table}[h!tbp]
\centering
\begin{tabular}{p{1.0cm} p{1.8cm}||c|c|c}
 \multicolumn{2}{c||}{\textbf{LTE networks}} & $ r_\mathrm{cell} $ & $ \sigma^\mathrm{A}_\mathrm{LTE} $ \cite{ryden_ltepos:15} & $ \sigma^\mathrm{B}_\mathrm{LTE} $ \cite{ryden_ltepos:15} \\
 \hline
 Italy& TIM      & 1.45 & 2.33 & 0.230 \\
      & Vodafone & 1.45 & 2.33 & 0.230 \\
      & Wind Tre & 1.25 & 2.04 & 0.200 \\
\multicolumn{2}{r||}{est. combined} & 0.79  & 1.28  & 0.130 \\
\hline
 Rome & TIM      & 0.33 & 0.53 & 0.053 \\
      & Vodafone & 0.26 & 0.42 & 0.042 \\
      & Wind Tre & 0.22 & 0.35 & 0.036 \\
\multicolumn{2}{r||}{est. combined} & 0.15 & 0.24 & 0.024 \\
\hline      
 Genoa& TIM      & 0.37 & 0.59 & 0.060  \\
      & Vodafone & 0.28 & 0.45 & 0.045  \\
      & Wind Tre & 0.66 & 1.10 & 0.105 \\    
\multicolumn{2}{r||}{est. combined} & 0.21 & 0.34 & 0.034 \\
\hline
\end{tabular}
\caption{Estimated individual positioning accuracies in scenario A and B for different LTE network providers in different territories, using the methods discussed in the text. All distances are given in \si{\kilo\meter}.}
\label{tab:lte_accuracies}
\end{table}

\subsection{GSM networks positional accuracies}

The first experiments of population density measurements in urban contexts were conducted about 15 years ago in Rome by the Senseable Cities group of the Massachusetts Institute of Technology (see \cite{ratti_rtrome:06}) using Telecom Italia Mobile (TIM) GSM towers and its Localizing and Handling Network Event System (LocHNESs) platform to measure in real-time people density and fluxes through the localization and mapping of each individual UE. Indeed, the location method implemented in the platform combined measured signal strengths from serving transceiver stations, their neighboring ones and from each mobile device to triangulate their locations every $ \sim\,\SI{0.5}{\second} $, in a similar way to what OTDOA does in the case of LTE networks. A spatial resolution of \SI{163}{\meter} was obtained in Rome's urban area with the available GSM network, increasing to \SI{295}{\meter} in the suburbs and to \SI{1235}{\meter} in the extra-urban area \cite{ratti_rturban:10}. More recent experiments adopting time difference of arrival localization algorithms have also shown to be able to break the \SI{50}{\meter} barrier in an urban context using GSM only \cite{laitinen_gsm:10}. Scaling the LocHNESs method resolution obtained in Rome by the GSM cells sizes in the different territories by these of Rome (see \fig{rome_hist}) allows to calculate the estimated resolution $ \sigma^\mathrm{B}_\mathrm{GSM} $ of GSM in a realistic scenario B. 
\par
The estimate of GSM resolution in a realistic scenario A (cell identity only) was obtained with the same method used for LTE, i.e. by considering the estimates in \cite{ryden_ltepos:15} to define $ \sigma^\mathrm{A}_\mathrm{GSM} \approx 1.6 \cdot r_\mathrm{GSM} $.
\par
A summary of the obtained positioning resolutions for the different contexts is given in \tab{gsm_accuracies}. Results in scenario B are comparable between GSM and LTE after being scaled by their different cell sizes, further suggesting that resolution is mostly set by geometrical disposition of the cells on the territory rather than technology-specific location protocols implementations. 

\begin{table}[h!tbp]
\centering
\begin{tabular}{p{1.0cm} p{1.8cm}||c|c|c}
 \multicolumn{2}{c||}{\textbf{GSM networks}} & $ r_\mathrm{GSM} $ & $ \sigma^\mathrm{A}_\mathrm{GSM} $ \cite{ryden_ltepos:15} & $ \sigma^\mathrm{B}_\mathrm{GSM} $ \cite{ratti_rturban:10} \\
 \hline
 Italy& TIM      & 2.10 & 3.36 & 0.357 \\
      & Vodafone & 2.11 & 3.37 & 0.359 \\
      & Wind Tre & 1.74 & 2.78 & 0.296 \\
\multicolumn{2}{r||}{est. combined} & 1.13 & 1.81 & 0.192 \\
\hline
 Rome & TIM      & 0.94 & 1.51 & 0.163 \\
      & Vodafone & 0.36 & 0.58 & 0.061 \\
      & Wind Tre & 0.70 & 1.12 & 0.120 \\
\multicolumn{2}{r||}{est. combined} & 0.30 & 0.48 & 0.051 \\
\hline      
 Genoa& TIM      & 0.34 & 0.54 & 0.058 \\
      & Vodafone & 0.32 & 0.51 & 0.054 \\
      & Wind Tre & -- & -- & -- \\    
\multicolumn{2}{r||}{est. combined} & 0.23 & 0.37 & 0.039 \\
\hline
\end{tabular}
\caption{Estimated individual positioning accuracies in scenario A and B for different GSM network providers in different territories, using the methods discussed in the text. All distances are given in \si{\kilo\meter}.}
\label{tab:gsm_accuracies}
\end{table}

\subsection{Reaching individual's proximity accuracy}

The spatial resolution analysis here performed shows that anonymous number of connections data (scenario A) could provide reliable information about population density on the scale of hundreds of meters in most urban contexts, i.e. allowing only big crowds to be distinguished, unless further capillarization of the networks are made available. Non-anonymous location data, on the other hand, would provide spatial resolutions in the order of the tens of meters (scenario B), i.e. the scale of small crowds and the individual proximity. 
\par
The studies reported before \cite{ryden_ltepos:15} underlined that significant improvements of the anonymous cell identity method can be obtained by deploying mini- and micro- cells inside buildings, enhancing the CI resolution to $ \ll \SI{100}{\meter} $. A further study from the same group \cite{ryden_ltepos2:16} showed also that indoor resolutions up to $ \sigma_\mathrm{OTDOA} = \SI{1.8}{\meter} $ can be reached by deploying indoor LTE pico-cells combined with an enhanced, iterative OTDOA approach. Finally, GPS-assisted UE locating (also known as A-GNSS), a part of the LTE standard, is also a solid possibility for resolution enhancement in outdoor contexts, reaching the scale of $ \approx \si{\meter} $. 
\par
Another possibility for the CI approach to reach the $ \sim \SI{10}{\meter} $ resolutions would be to complement mobile location data with higher resolution local data sources, such as public WiFi hotspots, public means of transportation locations, smart mobility devices locations, RF-noise levels monitoring tools. As these networks are much less homogeneous than mobile network between different urban contexts, calibration of the reconstructed density maps with mobile network devices would likely be necessary anyway. To date, no known protocols are available for individual users to contribute with their own access points to an hypothetical WiFi collective monitoring network. 
\par
Levels of resolution in the \si{\meter} range would allow social distancing regulations to be enforceable at the individual scale, posing however the significant challenge of balancing the risks to individual privacy loss (see Privacy section). Practical use cases in means of public transportation, individual businesses, industries, offices and marketplaces would become a solid possibility, requiring only further capillarization of mobile network technologies - a trend that is already happening with the development of 5G networks. A summary of the identified methods per resolution provided is shown in \tab{accuracy_summary}.

\begin{table}[h!tbp]
\centering
\begin{tabular}{p{5.0 cm} p{3.0cm}}
 \multicolumn{2}{c}{\textbf{Available positional methods}} \\
 \hline
 $ \approx \SI{100}{\meter} $ (interact. w. area) & Cell Identity \\
 $ \approx \SI{10}{\meter} $ (interact. w. proximity) & OTDOA, Triang. \\
 $ \approx \SI{1}{\meter} $ (interact. w. individuals) & A-GNSS (outdoor),   OTDOA+ (indoor) \\ 
\hline
\end{tabular}
\caption{Summary of available mobile-network-based positioning methods per spatial resolution scale.}
\label{tab:accuracy_summary}
\end{table}

\section{Privacy}

The possibilities offered by existing technologies in mass locating people with spatial accuracies at the level of single individuals proximity, discussed in the above sections, raise privacy concerns that, if not properly handled, would prevent the adoption of a mass RTLS as the one here proposed.
\par
On the one hand, as stated by Fisher and Dobson in \cite{fisher_dobson:03}, \textit{``each individual should be able to negotiate access by another person to information about their location. No one else should be able to circumvent that right''}. On the other hand, as discussed by Ratti \textit{et al.} in \cite{ratti_mobileland:06}, there is general consensus that aggregated population density maps can be made to satisfy all requirements of data anonymity, i.e. without violating any personal data protection regulation. Fisher and Dobson also point out in \cite{fisher_dobson:03} that as long as personal data is not made available by mobile phone operators to third parties, most privacy concerns are avoided. 
\par
Here we first review the juridical literature on personal data privacy relevant for this matter, i.e. that defining law boundaries to data anonymity. Then we take a quantitative approach to address the implementation of the privacy constraints, laying down the specifications of the second key part of the system: the user data anonymization and mapping algorithm. 


\subsection{Juridical boundaries}

The boundaries of data anonymity are sketched by recital 26 of Regulation (EU) 2016/679 of the European Parliament and of the Council of 27 April 2016 on the protection of natural persons with regard to the processing of personal data and on the free movement of such data, and repealing Directive 95/46/EC (General Data Protection Regulation) (see \cite{eu_GDPR:16}), stating that: \textit{``The principles of data protection should therefore not apply to anonymous information, namely information which does not relate to an identified or identifiable natural person or personal data rendered anonymous in such a manner that the data subject is not or no longer identifiable''}. The boundaries of identifiability are also sketched in the same recital 26: \textit{``To ascertain whether means are reasonably likely to be used to identify the natural person, account should be taken of all objective factors, such as the costs of and the amount of time required for identification, taking into consideration the available technology at the time of the processing and technological developments''}. This formulation implies that technical analysis to evaluate the feasibility and the efforts required to identify the natural person to which data relate have to be carried out during the design phase of any information systems involving personal data, leaving open the data anonymization modality. 
\par{}
Aggregated data that can be proven to fall under these prescriptions can thus be declared anonymous; for these, the recital clearly states: \textit{``(the General Data Protection Regulation) does not concern the processing of such anonymous information, including for statistical or research purposes''}. Thus, if proper mitigation of the risk that a single person's identity can be retrieved is carried out, real-time population density maps can be considered anonymous data and thus publicly distributed.

\subsection{Ensuring data anonymity}

A standard adopted technique for anonymizing mapping data consists in introducing a minimal threshold on the number of individuals per cell, under which the cell is discarded from the map. This guarantees a high combinatorial cost for following single individuals on the map. For instance, if an average of 4 individuals per map cell is present, in order to reconstruct the trajectory of one of them starting from a given cell in the map, the whole combinatorial tree of possible steps into nearby valid cells should be considered, whose number of branches grows by $ 4^n $ where $ n $ being the number of steps. This is the principle beyond anonymizing by thresholding: making de-facto impossible the reconstruction of a single trajectory, and thus its de-anonimization, due to the very high number of equally-likely possible combinations.
\par
This method has a drawback: not all of the available information is used to generate the map, as that from a certain fraction of the individuals is discarded after thresholding. The amount of information lost is by construction highly non-linear in the number of people per cell, causing critical statistics losses when the size of the mapping cells is small enough for the average number of people per cell to be comparable or smaller than unity. In other words, the choice of the thresholding level imposes a minimal resolution to the mapping system, under which most of the information available is discarded.
\par
In the present case, this drawback would significantly limit the maximal resolution of the proposed system and its risk-prevention capability, as its goal is by construction to get as close as possible to the single individual scale.

\subsection{Anonymizing positions preserving interdistances}

To avoid this thresholding inconvenience, we developed a position anonymizing system working independently from the spatial resolution scale choosen for mapping. It is based on preserving the key information necessary for prevention, i.e. mutual interdistance between individuals, with the highest accuracy, while anonymizing individuals' absolute positions.
\par
Here for sake of simplicity we describe the approach in a mono-dimensional scenario; the bi- and tri-dimensional cases follow as its straightforward generalizations.

\begin{figure*}[h!tbp]
	\centering		
    \includegraphics[width=\linewidth]{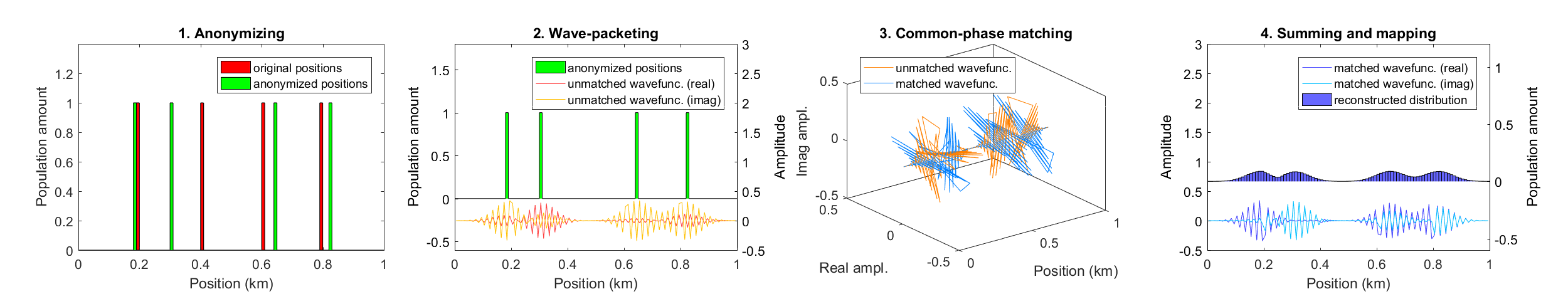}
    \includegraphics[width=\linewidth]{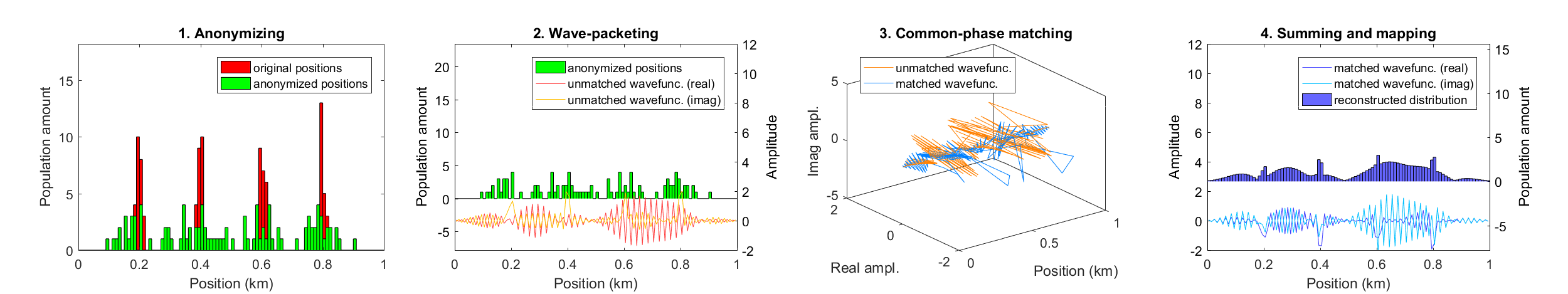}
    \includegraphics[width=\linewidth]{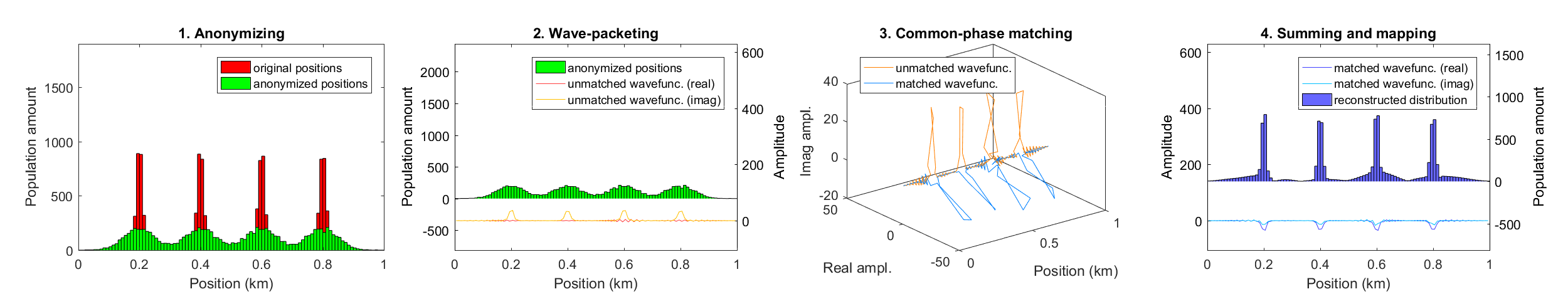}  
	\caption{Anonymization and privacing example in a mono-dimensional domain. The columns refer to the four steps of the anonimyzing and mapping algorithm described in the text. The three rows are obtained with different number of individuals: $ N = 4$ (top), $ N = 100 $ (middle) and $ N = 10000 $ respectively. }
	\label{fig:privacing1}
\end{figure*}

\begin{enumerate}
\item \textbf{Anonymizing:} the first step (see \fig{privacing1}.1) consists in collecting from the localization system the set of \textit{original positions} $ {x_0}^{(n)} $ (here supposed exact) of each of the $ N $ individuals in the mapping domain ($ n \in [1,N] $). A set of random positions $ {x_1}^{(n)} $ are then generated following Gaussian distributions $ G_{[{x_1}^{(n)}, \sigma^{(n)}]}(x) $ centered on $ {x_0}^{(n)} $ with standard deviations $ \sigma^{(n)} $. These generated positions $ {x_1}^{(n)} $ take the name of \textit{anonymized position}, whereas the coefficients $ \sigma^{(n)} $ take the name of \textit{anonymization parameters}.
\item \textbf{Wave-packeting:} the second step (see \fig{privacing1}.2) consists in constructing \textit{wave packets} $ \psi^{(n)}(x) $ relative to each anonymized position $ {x_1}^{(n)} $. Each $ \psi^{(n)}(x) $ is a complex-valued function with complex-amplitude given by the same Gaussian distribution used for generating anonymized positions, $ G_{[{x_1}^{(n)}, \sigma^{(n)}]}(x) $, and complex-phase dependent upon the distance between each point of the grid $ x $ and the original individual position $ {x_0}^{(n)} $, $ \exp(i k || {x_0}^{(n)} - x||) $, where $ k = \pi M / L $ is the \textit{spatial sampling wavenumber}, here chosen (following Nyquist theorem) for a full wavelength to match $ 2 $ times the map resolution $ d = L / M$ ($ L $ being the whole map size). All wave packets are summed up in a collective \textit{wave-function} $ \psi(x) = \sum_{n=1}^N \psi^{(n)}(x) $, which is then sampled over the $ M $ points $ p^{(m)} $ of the regular grid chosen for mapping (here $ m \in [1,M] $) to build the aggregated complex-valued map $ \psi^{(m)} \equiv \psi(p^{(m)}) $. $ \psi^{(m)} $ takes the name of \textit{unmatched wavefunction} since its phase reference is still arbitrary. It is no longer possible to reconstruct the original positions $ {x_0}^{(n)} $ from $ \psi^{(m)} $ to better than $ \sigma^{(n)} $, as each of the individual wave-functions are centered on the Gaussianly-randomized positions $ {x_1}^{(n)} $. Whether this aggredated map can now be considered anonymous or not, it depends only on the choice of $ \sigma^{(n)} $ rather than that of the mapping accuracy. If $ \sigma^{(n)} $ coefficients in the anonymization step are chosen so that many individuals' wave-packets overlap with each other, combinatorial anonymization is again provided (see next sections) while surpassing the limitation of loosing information of anonymization by thresholding. This process thus allows constructing aggregated, anonymized complex-valued unmached wavefunctions from $ H $ localization system providers, $ \psi_h^{(m)} $ (with $ h \in [1,H] $), that can now be individually distributed to a third-party.
\item \textbf{Common-phase matching:} the third step (see \fig{privacing1}.3) is run by the system collecting all individual unmatched wavefunctions $ \psi_h^{(m)} $ received from the individual providers. A phase rotation $ \exp(i \phi_h) $ is applied to each of the unmatched wavefunction, allowing each one to be referenced to the same phase-referencing system and thus constructing \textit{phase-matched wavefunctions} that can be now summed up constructively. Unmatched wavefunctions are assumed here to be built on the same previously-set mapping grid $ p^{(m)} $. Phase rotation coefficients allow, in other words, all individual maps to be cross-calibrated on the same positional referencing system.
\item \textbf{Summing and mapping:} the fourth an last step (see \fig{privacing1}.4) consists in summing all matched wavefunctions together and then take the complex-modulo operation to build the final population density map $ \rho^{(m)} = | \sum_{h=1}^H \exp(i \phi_h) \psi_h^{(m)}| $. This obtained density map can be shown to be equal to the exact distribution of people (i.e. that built with the original positions $ {x_0}^{(n)} $) in the limit of high $ N $, whereas in the limit of low $ N $, it is set mostly by the anonymized positions $ {x_1}^{(n)} $ (see next section).
\end{enumerate}

It can be shown that the process here described alters the positions of the single individuals while maintaining the possibility to calculate their mutual interdistances, encoded as complex-phase differences. The working principle of this method is easily understood if one computes the collective wave-function $ \psi(x) $ for a simplified case in which only two-individuals are present. Here $ \psi(x) $ takes the form:

\begin{align*}
\psi & = G_1 \, e^{i k (p^{(m)} - {x_0}^{(1)})} \, + \, G_2 \, e^{i k (p^{(m)} - {x_0}^{(2)})}\\
\end{align*}

where, for notation simplicity, we set $ G_1 \equiv G_{[{x_1}^{(1)}, \sigma^{(1)}]}(x) $ and $ G_2 \equiv G_{[{x_1}^{(2)}, \sigma^{(2)}]}(x) $. Taking the complex squared-modulo, one gets:

\begin{align*}
|\psi|^2 = {G_1}^2 + {G_2}^2 + 2 \, G_1 \, G_2 \cos (k ({x_0}^{(1)}-{x_0}^{(2)})) \\
\end{align*}

The two Gaussian terms $ G^{(1)}(x) $ and $ G^{(2)}(x) $ are centered in $ {x_1}^{(1)} $ and $ {x_1}^{(2)} $, thus not allowing position de-anonimizing from a single map, while their interference contains the oscillatory term $ \cos (k ({x_0}^{(1)}-{x_0}^{(2)})) $ which encodes in the oscillations the reciprocal distance between the individuals, referenced to the mapping grid and calculated with their original positions. The maximum amount of information that can be thus retrieved from the final map about the two individuals are the two anonymized positions, their mutual inter-distance and the expected area in which each of the individuals is with a given probability. 
\par
This process thus protects individual privacy in the limit of a few people present in the map, an example of which is shown in \fig{privacing1}, first row, in which 4 people are positioned at regular \SI{200}{\meter} distances in a \SI{1}{\kilo\meter} context being mapped with \SI{10}{\meter} resolution (\fig{privacing1}, top left). The final population density distribution produced by the algorithm shows four Gaussian functions centered on the anonymized positions and delocalized with $ \sigma = \SI{50}{\meter} $. When the number of individuals grows to 100 (\fig{privacing1}, second row) and 10000 (\fig{privacing1}, third row), progressively all interference terms sum in phase and ultimately dominate over the stochastic Gaussian terms, giving rise again to the original density distribution in the statistical limit of many people (compare \fig{privacing1} bottom row, red to blue distributions). In all these cases it is always impossible from the final distribution to retrieve the exact original positions, as the first stochastic anonymization cannot be inverted. 
\par
It's worth noticing that this inversion cannot be performed exactly even considering many different realizations of the maps at different time instants, as the stochastic anonymizing process give different anonymized positions at each time. This task can be shown to be equivalent to retrieving the exact position of a quantum particle at rest obeying Heisenberg's uncertainty principle. Its exact position would be found only in the limit of infinite observations, and only if the particle is at rest all the time (i.e. with null average velocity). Combining many observations over a long time of a well isolated single individual in quiet may still allow to reduce the positional uncertainty statistically down to an undesired high level of accuracy. 
Several approaches are being investigated to prevent this from happening by design. The simplest one is the addition of a small amount of white noise to each of the complex maps. A more sophisticated approach is to introduce some combinatorial anonymization, obtainable in this context (without any loss of information) by selecting large enough $ \sigma^{(n)} $ for several wave-packets to overlap with each other in any point of the map. Here the challenge is to select the values of $ \sigma^{(n)} $ following the local availability of statistics and adopting an optimal resolution-to-anonymization criterion. This approach is currently being investigated.

\begin{figure*}[h!tbp]
	\centering		
    \includegraphics[width=\linewidth]{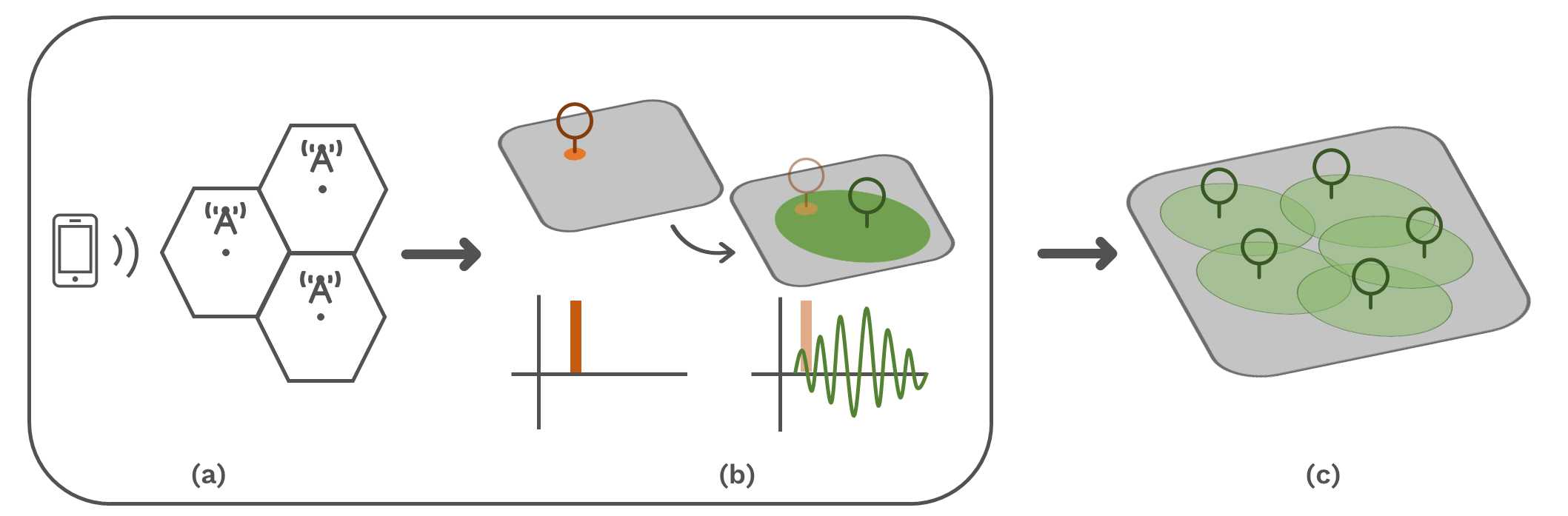} \\
    \vspace{1cm}
    \includegraphics[width=\linewidth]{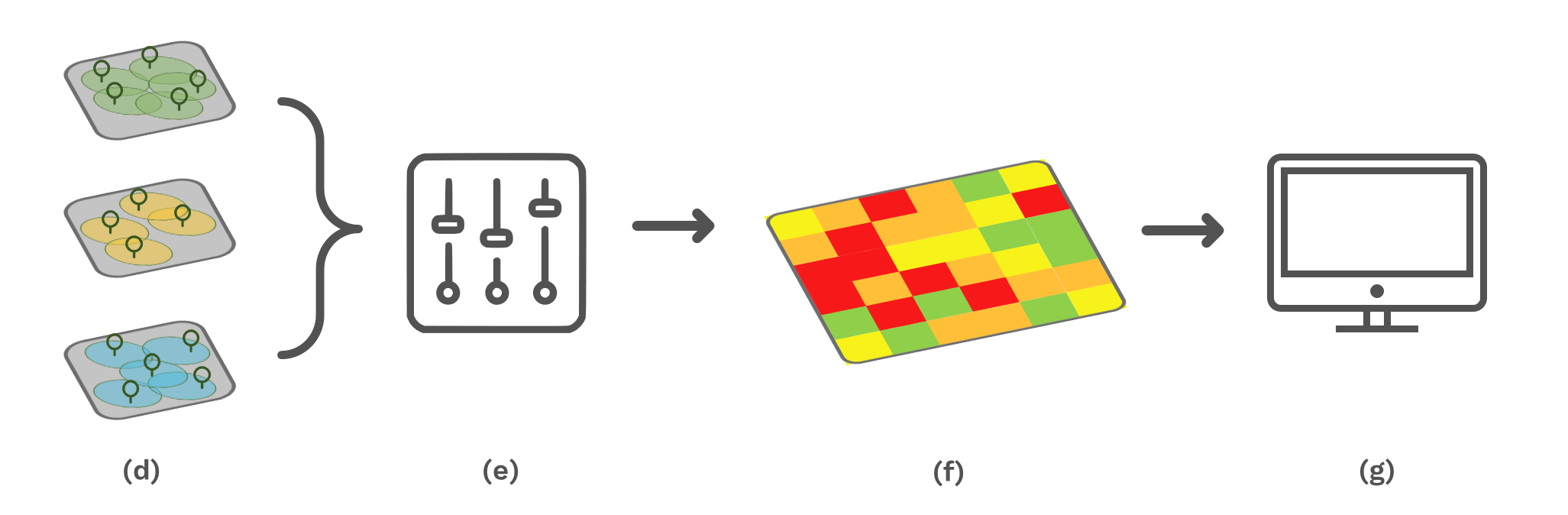}
	\caption{System overview: the first part of the data processing is under the responsibility of each location data provider (gray box): (a) individual users positions are detected and collected by the network provider data system; (b) the anonymization algorithm is run within the location data provider computing infrastructure in order to delocalize each users' position into quantum-like complex-valued wave packets; (c) all wave packets are combined into an aggregated, anonymized wave function; (d) several wave functions are obtained from different mobile network system providers and are distributed to a third-party for subsequent analyses; (e) all providers data are calibrated and combined in order to obtain the final density map, anonymized yet faithful to the real people density (f) that can be published and diffused widely among stakeholders (g).}
	\label{fig:system_overview}
\end{figure*}

\section{A possible system overview}

A sketch of the proposed data processing system is shown in \fig{system_overview}. The system is composed of two main parts: a first data processing pipeline hosted in each data provider computing facility and a second anonymous data aggregation and post-processing pipeline hosted in a common computing infrastructure providing also open data publication. 
\par
The first part of the data processing falls inevitably under the responsibility of each location data provider, as they must comply with the strict policies about personal data handling. This processing involves the collection of users' locations and their subsequent anonymization into a single anonymous, aggregated complex-valued map, following the algorithm described in the previous sections. Each of these maps, being now anonymous and compliant with privacy regulations (also discussed in the previous sections), can now be transmitted to the common computing infrastructure that collects all of them from the different location providers participating in the initiative. 
\par
The role of the common project infrastructure is to run the common-phase matching and mapping algorithms also discussed in the previous sections, to produce the final population density maps which are made available openly. The frequency of new maps generation has to lie in the range of minutes, in order to be effective for prevention on typical human time scales.

\section{Conclusions}

This paper proposes the start of an initiative for developing a privacy-protecting real-time, open-data population density and flux mapping system, hypothetically named \textit{PeopleTraffic}, designed to support citizens and public authorities in harmonizing epidemic diffusion risks whilst properly dealing with privacy concerns. In the specific, it is meant as a tool for aiding the management of the COVID-19 pandemics Phase-2, which is expected to last approximately until either population herd immunity is obtained or a vaccination becomes available to the mass public.
\par
The proposed system is based upon recognizing that social distancing regulations, enforcing physical separation between individuals to limit population density and reducing individual-to-individual contacts, have been the most effective tool in mitigating the spread of epidemics driven by individual-to-individual contact (as COVID-19 and most influenza viruses). 
\par
The proposed system makes use of the nowadays capillary distributions of mobile network stations (i.e. these adopting GSM, UMTS, and LTE network technologies) to construct real-time population density and flux maps with sufficient spatial and temporal resolutions to get close to the individual proximity scale. The construction of these maps includes an innovative approach to privacy-by-design, compliant with the current European privacy regulations which, thanks to a quantum-inspired delocalization process, preserves the key information necessary for prevention (i.e. mutual inter-distance between individuals) with the highest accuracy while anonymizing individuals' absolute positions. The proposed method is robust against incomplete cell phone possession or activation, intentional de-activation of cell phones by their owners or prevalence of specific providers in specific areas.
\par
In place of ex-post intervening when positive cases of infection are detected (which typically happens with some time-lag from the diffusion event), the here-obtained real-time people distribution maps support preventively orienting people towards safe behaviour allowing, e.g., choosing the timing and the means of transportation for commuting. They would also provide valuable information to policy-makers in taking decisions based on people density, which has proved so far one of the most reliable a-priori predictors of infection spreading. 

\section{Acknowledgments}
The work beyond this note was conducted during the first outbreak of the COVID-19 pandemic in 2020. Despite the lockdown, I kept receiving funding from the European Union’s Horizon 2020 research and innovation programme under the Marie Skłodowska-Curie grant agreement No 754496 FELLINI, which I want to acknowledge here. I also wish to thank all that contributed directly to this whitepaper, with the important input and discussions from:

\begin{itemize}[noitemsep,topsep=0pt,parsep=0pt,partopsep=0pt]
    \item Dr. Alessandra Casale (Rulex Innovation Labs.);
    \item Prof. Fiorella De Cindio (Dept. of Computer Science, Universit\`a Statale di Milano and Fondazione RCM) and Mr. Giuseppe Caravita (Il Sole 24 Ore, retired);
    \item Dr. Michael Doser (Experimental Physics dept., CERN) and Mrs. Silvia Wyder (PhD cand., Dept. Art Therapy and Cultural Studies, University of Derby, UK);    
    \item Avv. Marco Ciurcina (Nexa Center for Internet \& Society and Politecnico di Torino)
    \item Dr. Christian Vincenot (Biosphere Informatics Lab., Dept. of Social Informatics, Kyoto University)   
	\item Prof. Stefano Mazzoleni (Applied Ecology and System Dynamics Lab., Universi\`a di Napoli 
``Federico II'')      
	\item Prof. Danilo Bruschi (Dept. of Computer Science, Universit\`a Statale di Milano)
	\item Prof. Elena Pagani (Dept. of Computer Science, Universit\`a Statale di Milano) 
	\item Prof. Alfonso Fuggetta (Politecnico di Milano and Cefriel)
	\item Dr. Marcello Scipioni (Dept. of Digital Innovation, Camera del Lavoro di Milano)            
\end{itemize}

\bibliographystyle{unsrt}
\bibliography{references}

\end{document}